\documentclass[%
 reprint,
 amsmath,amssymb,
 aps,
 prl,
]{revtex4-1}
\usepackage{float}
\usepackage{graphicx}
\usepackage{dcolumn}
\usepackage{bm}

\begin{document}

\preprint{APS/123-QED}

\title{Transverse Kerker effect of localized electromagnetic sources}

\author{Feifei Qin$^{1}$, Zhanyuan Zhang$^{1}$, Kanpei Zheng$^{1}$, Yi Xu$^{1,2}$}
\email{yi.xu@osamember.org}
\author{Songnian Fu$^{2}$}
\author{Yuncai Wang$^{2}$}
\author{Yuwen Qin$^{2}$}

\affiliation{$^{1}$Department of Electronic Engineering, College of Information Science and Technology, Jinan University, Guangzhou, 510632, China}
\affiliation{$^{2}$Advanced Institute of Photonics Technology, School of Information Engineering, and Guangdong Provincial Key Laboratory of Information Photonics Technology, Guangdong University of Technology, Guangzhou, 51006, China}

\date{\today}

\begin{abstract}
Transverse Kerker effect is known by the directional scattering of an electromagnetic plane wave perpendicular to the propagation direction with nearly suppression of both forward and backward scattering. Compared with plane waves, localized electromagnetic emitters are more general sources in modern nanophotonics. As a typical example, manipulating the emission direction of a quantum dot is of virtue importance for the investigation of on-chip quantum optics and quantum information processing. Herein, we introduce the concept of \textit{transverse Kerker effect of localized electromagnetic sources} utilizing a subwavelength dielectric antenna, where the radiative power of magnetic, electric and more general chiral dipole emitters can be dominantly directed along its dipole moment with nearly suppression of radiation perpendicular to the dipole moments. Such transverse Kerker effect is also associated with Purcell enhancement mediated by electromagnetic multipolar resonances induced in the dielectric antenna. Analytical conditions of transverse Kerker effect are derived for the magnetic dipole, electric dipole and chiral dipole emitters. We further provide microwave experiment validation for the magnetic dipole emitter. Our results provide new physical mechanisms to manipulate the emission properties of localized electromagnetic source which might facilitate the on-chip quantum optics and beyond.

\end{abstract}

\maketitle



\textit{Introduction.---}
Electromagnetic scattering of subwavelength particles is of broad interesting because its vital importance in understanding the fundamental science of various phenomena in our daily life \cite{Jackson,Bohren}, such as the blue sea and sky. Kerker effect, manifested itself as giant asymmetry of forward-to-backward electromagnetic scattering, is a simple while universal physical mechanism to direct the scattering of an electromagnetic plane wave utilizing Mie resonant particles \cite{Kerker}. In particular, an ultra-high ratio of forward-to-backward can be achieved in the first Kerker condition or vice versa in the second Kerker condition for the backward-to-forward scattering \cite{Kerker,NC}. The generality of this canonical unidirectional scattering mechanism relies on the interference of electromagnetic multipolar moments induced in the Mie scatters \cite{Kerker,NC,directional1,directional2,directional3,directional4,directional5,directional6,directional7,directional8,directional9,directional10,wei1,wei2}. Distinct from the Kerker conditions of unidirectional scattering parallel to the propagation direction of an electromagnetic plane wave, transverse Kerker effect enables the scattering perpendicular to the propagation direction with simultaneous suppression of forward and backward scattering \cite{transverse1,transverse2}. Both longitudinal and transverse type of Kerker scattering provide unified mechanisms for molding the flow of an ideal plane wave that is the simplest solution of Maxwell's equation in the homogeneous medium. However, localized emitters are more general electromagnetic sources in practice, representative examples are including but not limit to quantum dots in quantum optics \cite{chiral,jin1,jin2} and wavelength scale antennas in communication \cite{antenna} etc. 

\begin{figure}[t!]
\includegraphics[width=\columnwidth]{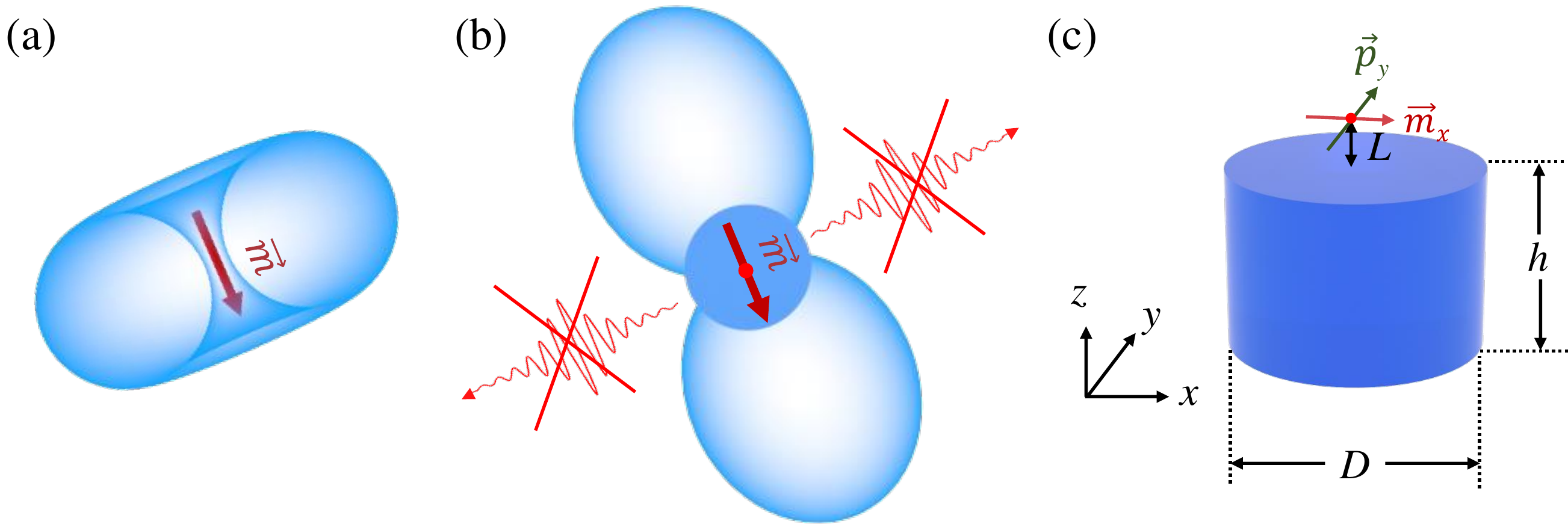}
\caption{\label{fig:fig2} 
(Color online)(a) Schematic of the far-field radiation power for a magnetic dipole emitter (red arrow) in a homogeneous medium. (b) By placing the dipole emitter close to a subwavelength dielectric antenna shown in (c), the far-field radiation power can be redirected along the direction of dipole moment while the radiation along the direction perpendicular to the dipole moment is substantially suppressed. The diameter $D$ and height $h$ of the cylinder are outlined in (c). The dipole emitter can be magnetic, electric or chiral dipole.
}
\end{figure}

Cavity quantum electrodynamics (QED) relies on the interaction between dipole type quantum emitters and a resonant cavity \cite{cavity}, enabling a various of attractive quantum-electrodynamics phenomena \cite{cqed}. In particular, the ability to manipulate the emission direction of dipole type quantum emitters \cite{chiral,jin1,jin2,chiral1} could pave the way for advanced integrated quantum physics \cite{wang,ren}. As a result, the investigation of unidirectional Kerker scattering of dipole emitters considering the directivity of either near- or far-field zone has received increasing research efforts these years \cite{chiral,jin1,near1,near2,chiral1,dipole1,dipole2,dipole22,dipole222,dipole3,dipole4,dipole5,dipole6,dipole7,dipole8}. In principle, the far-field emission power along the direction of dipole moment is zero, as schematically shown in Fig. 1 (a). By interacting the dipole emitter with subwavelength particles of various morphologies, there is possibility to direct the emission parallel to the direction of dipole moment \cite{dipole1,dipole22,dipole3,dipole4,dipole5}, analogue to its transverse Kerker effects of plane wave counterpart demonstrated recently \cite{transverse2}. However, the mechanisms proposed in these pioneer works \cite{dipole1,dipole22,dipole3,dipole4,dipole5} seems to be not universal and no unified theory for the transverse Kerker conditions of localized electromagnetic source has been proposed. Therefore, the quest to the general physics for transverse Kerker effect of localized electromagnetic source, which would become an essential knob of integrated quantum optics, remains an alluring but unrevealed issue.

In this letter, we pinpoint a concept of \textit{transverse Kerker effect of localized electromagnetic sources} and provide a general theory for the transverse Kerker condition for various kinds of dipole emitters, including magnetic dipole (MD), electric dipole (ED) and chiral MD/ED emitters. Microwave experiment of MD emitter is provided to further consolidate our theory.

\textit{Theory and simulation.---}
Without loss of generality, we consider a typical localized electromagnetic source, i.e. a dipole type emitter, in the vicinity of ($L \ll \lambda$) a subwavelength dielectric antenna supporting electromagnetic multipolar resonances \cite{science}, as shown in Fig. 1 (b) and (c). We consider the situation that the retardation effect of this system can be neglected \cite{ja1}, which resembles the realistic condition that a quantum dot embedded in a resonant nanoantenna \cite{jin1}. In general, the dipole emitter will polarize the antenna where various electromagnetic multipole moments of the antenna will be induced. Following Jackson's formulation \cite{Jackson}, the properties of localized electromagnetic sources in the vicinity of a dielectric antenna can be characterized by a series of spherical harmonics with the coefficients $a_E(l,m)$ and $a_M(l,m)$ specifying the amounts of electric and magnetic multipole fields, respectively (see Notes. 1 of the Supplemental Material \cite{Si}). Therefore, the expression for the angular distribution of radiated power can be written as follows:
\begin{eqnarray}
\begin{aligned}
\label{eq:eq2}
\frac{dP(\theta,\phi)}{d\Omega}=\frac{Z_{0}}{2k^2}\bigg|\sum_{l,m}(-i)^{l+1}[a_{E}(l,m)\textbf{X}_{lm}\times\textbf{n}\\+a_{M}(l,m)\textbf{X}_{lm}]\bigg|^2,
\end{aligned}
\end{eqnarray}
where $\textbf{n}$ = $\textbf{r}/r$ is the unit vector from the origin to the observation point, $(r,\theta,\phi)$ are the spherical coordinates of $\textbf{r}$, $Z_0$ is the impedance of vacuum, $k$ is the wavenumber, and $\textbf{X}_{lm}$ = $1/\sqrt{l(l+1)}\textbf{L}Y_{lm}$ [$\textbf{L}$ is the orbital angular momentum operator and $Y_{lm}$ are spherical harmonics of order($l>0$, $0\leqslant|m|\leqslant l)$]. The multipolar coefficients associated to spherical harmonics at the right hand side of Eq. (1) is closely related to the combined symmetry of the localized source and antenna, which means that coefficients with specified $l,m$ in Eq. (1) could vanish because of symmetry (see Notes. 2 in the Supplemental Material \cite{Si}). 

As an illustrative example, we consider the scenario of a MD (ED) emitter along $x$ axis ($y$ axis) interacts with a cylindrical antenna with $D_{\infty h}$ symmetry where the emitter locates at the symmetry axis of the antenna with a horizontal dipole moment [see Fig. 1 (c)]. According to the symmetry of each component for the electromagnetic field which determines the projection to a specified spherical harmonic, one can obtain $a_M(l,m)$ = $(-1)^{m}a_M(l,-m)$, $a_E(l,m)$ = $(-1)^{m+1}a_E(l,-m)$ and $a_{E(M)}(2,\pm2)$ = 0, respectively, where the contributions of high order multipole fields are neglected here ($l\leqslant2$). While for the MD (ED) emitter along $y$ axis ($x$ axis), the situation changes to $a_M(l,m)$ = $(-1)^{m+1}a_M(l,-m)$ and $a_E(l,m)$ = $(-1)^{m}a_E(l,-m)$. For the subwavelength dielectric antenna shown in Fig. 1 (c), the remaining coefficients are $a_{E(M)}(1,\pm 1)$ and $a_{E(M)}(2,\pm 1)$. The transverse Kerker condition for the aforementioned MD or ED emitter can be derived as (see Notes. 2 for the general expression in the Supplemental Material \cite{Si}): 

\begin{eqnarray}
\label{eq:eq2}
\frac{dP^T}{d\Omega}(\theta=180^{o})=\frac{Z_0|A+iB|^2}{8{\pi}k^2}=0,
\end{eqnarray}
\begin{eqnarray}
\label{eq:eq3}
\frac{dP^T}{d\Omega}(\theta=0^{o})=\frac{Z_0|A-iB|^2}{8{\pi}k^2}=0,
\end{eqnarray}
where $A$ = $\sqrt{3}a_{E}(1,1)-\sqrt{5}a_{M}(2,1)$ and $B$ = $\sqrt{3}a_{M}(1,1)+\sqrt{5}a_{E}(2,1)$. Equation. (2) corresponds to the suppression of the radiation along $-z$ direction and Eq. (3) corresponds to the suppression of the radiation along $+z$ direction, which indicates that there is possibility to tailor the electric and magnetic multipole fields to achieve the transverse Kerker effect of MD or ED emitter.

To gain further insights into the underlying physics of the transverse Kerker scattering of a dipole emitter, we first examine a MD emitter (along $x$ axis) locates in close proximity to a dielectric antenna with a height $h$ = $\lambda_{0}/3.95$ and diameter $D$ = 2$r$ = $\lambda_{0}/2.1$, where $\lambda_{0}$ is the free-space wavelength at the transverse Kerker condition, as shown in Fig. 1(c). The distance between the MD and the upper surface of the antenna is $L$. The theoretical results obtained from Eq. (1) (solid lines) are shown in Fig. 2 (a), where we adopt the numerically extracted multipolar coefficients of $a_E(l,m)$ and $a_M(l,m)$ ($0<l\leqslant2$) by the finite element method (FEM). The conditions of Eq. (2) and (3) can be matched at the size parameter $kr$ of 1.49, where the backward and forward scattering approach to 0 simultaneously. Furthermore, the good agreement between theoretical and numerical (symbols) results in Fig. 2(a) also indicates that the expansion order of multipolar coefficient ($l$) up to 2 is sufficient. To further provide an intuitive picture, the far-field radiation patterns evaluated by Eq. (1) at the size parameter $kr$ of 1.49 are presented in Fig. 2(b). As can be seen from the angular distribution of radiated power, the scattering along the forward and backward directions are almost suppressed. At the same time, the dominant radiative direction (red solid line) is rotated about 90$^{\circ}$ compared with the result in vacuum (blue dashed line), indicating the transverse Kerker condition is achieved for a MD emitter. The FEM results are also provided in Fig. 2 (b) for comparison, where qualitative agreement between theoretical and numerical results can also be obtained. Furthermore, the enhancement of spontaneous emission can be also achieved by comparing the far-field scattering patterns of the dipole emitter with or without the antenna shown in Fig. 2 (b). The Purcell factor (PF) \cite{Purcell1,Purcell2,Purcell3,Purcell4}, i. e., the ratio of spontaneous decay rates with ($\gamma$) or without ($\gamma_{0}$) the structures is about 5 at the transverse Kerker condition. Such Purcell enhancement can be attributed to the coupling between the MD emitter and electromagnetic multipole modes of dielectric antenna, as validated by the distributions of electromagnetic field shown in Fig. 2(e). 

\begin{figure}[!htb]
\includegraphics[width=\columnwidth]{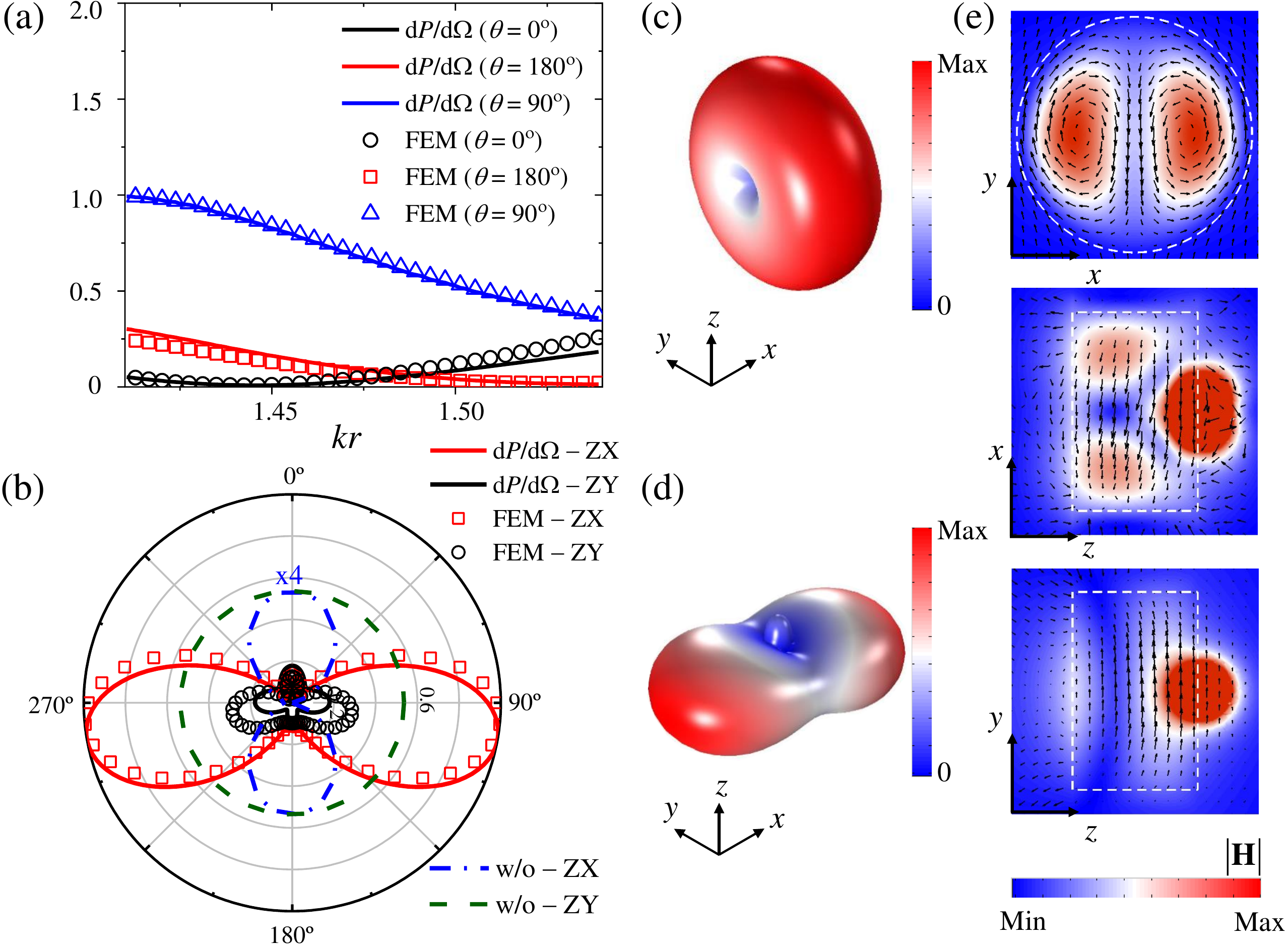}
\caption{\label{fig:fig2} 
(Color online)(a) Normalized time-averaged radiated power as functions of the size parameter $kr$ along three specified directions ($\theta$ = 0$^{\circ}$, 90$^{\circ}$ and 180$^{\circ}$) when a MD emitter is placed close to the dielectric antenna. The MD emitter (along $x$ axis) locates above the surface of the cylinder. Here, $\theta$ = $0^{\circ}$ indicates the radiation along the $+z$ axis while $\theta$ = $180^{\circ}$ refers to the radiation along the $–z$-direction. The case of $\theta$ = $90^{\circ}$ indicates the direction of transverse scattering paralleled to $x$-$y$ plane. (b) Two-dimensional transverse far-field scattering patterns at the size parameter of $kr$ = 1.49. Analytical results and numerical results are outlined by solid lines and symbol lines, respectively. Three dimensional far-field radiation pattern without and with the resonator are presented in (c) and (d), respectively. The magnetic field distribution at the transverse Kerker condition on the $xy$, $xz$ and $yz$ planes are shown in (e). Arrows indicate the distribution of vectorial electric field.
}
\end{figure}

\begin{figure}[!htb]
\includegraphics[width=\columnwidth]{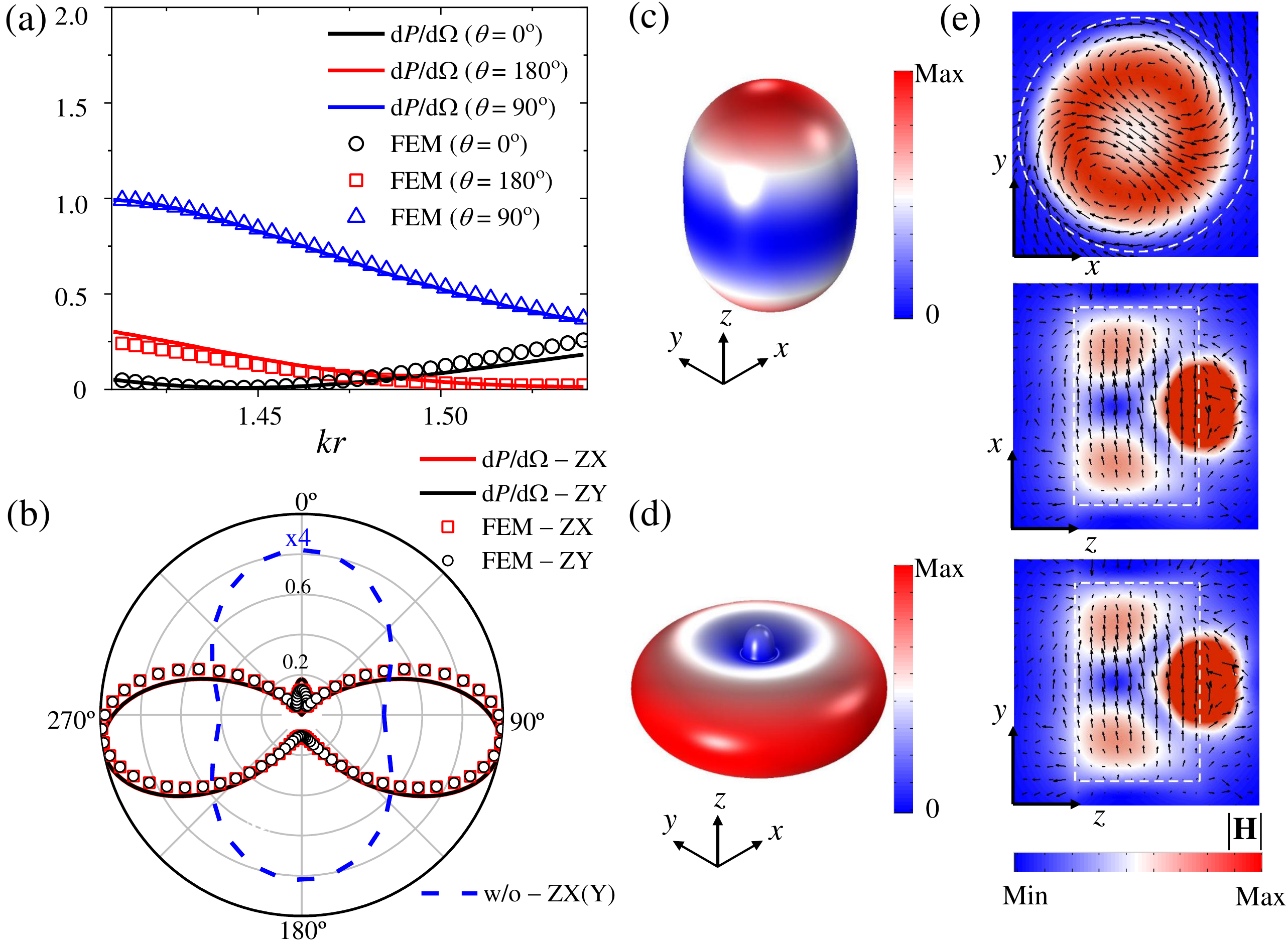}
\caption{\label{fig:fig3} 
(Color online)(a) Normalized time-averaged radiated power as functions of the size parameter $kr$ along three specified directions ($\theta$ = 0$^{\circ}$, 90$^{\circ}$ and 180$^{\circ}$) when a chiral MD emitter is placed closed to the dielectric antenna. The chiral MD emitter ($\vec{m}_{x}$+$i\vec{m}_{y}$) locates above the surface of the cylinder. (b) Two-dimensional transverse far-field scattering patterns at the size parameter of $kr$ = 1.49. Analytical results and numerical results are outlined by solid lines and symbols, respectively. Three dimensional far-field radiation pattern without and with the resonator are presented in (c) and (d), respectively. The magnetic field distribution at the transverse Kerker condition on the $x-y$, $x-z$ and $y-z$ planes are shown in (e). Arrows indicate the distribution of vectorial electric field.
}
\end{figure}

The transverse Kerker effect can also be generalized to the chiral MD emitter (see Notes. 2 in the Supplemental Material \cite{Si}). Figure 3 (a) presents the normalized time-averaged radiated power along three specified directions ($\theta$ = 0$^{\circ}$, 90$^{\circ}$ and 180$^{\circ}$), where the same dielectric antenna of Fig. 2 is excited by a chiral MD emitter ($\vec{m_{x}}$+$i\vec{m_{y}}$) at the same position. As can be seen from this figure, the transverse Kerker condition for the chiral MD emitter can also be achieved around the size parameter $kr$ of 1.49. The analytical and simulation results of the far-field scattering patterns shown in Fig. 3 (b) also demonstrate that the dominant radiation direction of the chiral MD emitter has been rotated by 90$^{\circ}$. According to 3D far-field scattering patterns shown in Fig. 3 (c) and (d) for the chiral MD emitter placed in vacuum and in the proximity of dielectric antenna, the homogeneous transverse Kerker scattering effect is more intuitively visualized. The predominant radiation along $z$ axis of a chiral MD emitter is isotropically redirected to the $x$-$y$ plane. The magnetic field distribution of the cylinder on the $x$-$y$, $x$-$z$ and $y$-$z$ planes are provided in Fig. 3 (e) at the transverse Kerker condition. Similarly, the enhancement of PF observed in Fig. 3 (c) also attributes to the electromagnetic multipolar resonances associated with strong concentration of electromagnetic field inside the antenna.

\begin{figure}[!htb]
\includegraphics[width=\columnwidth]{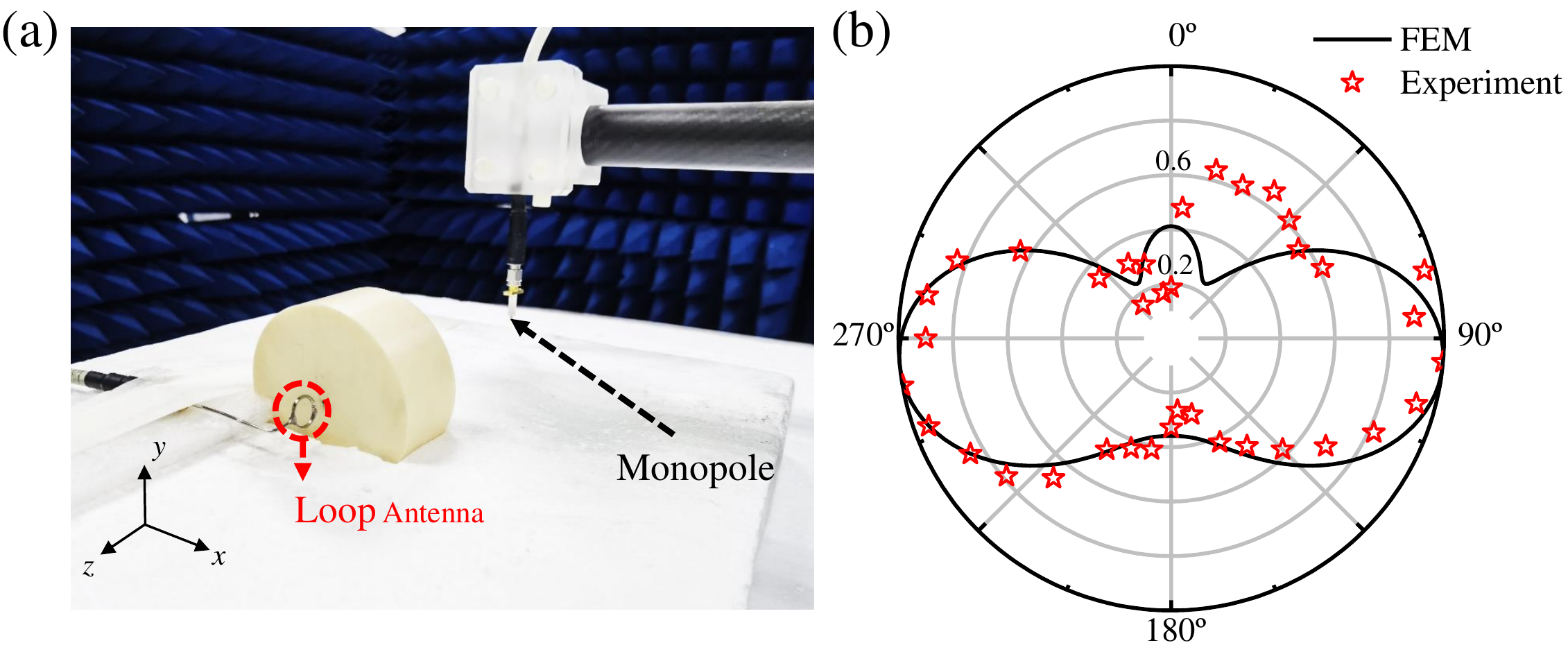}
\caption{\label{fig:fig4} 
(Color online) (a) Experimental setup to measure the far-field radiative patterns. (b) Radiative patterns at the size parameter of $kr$ = 3.04, where the black solid lines corresponds to the numerical results on $x-z$ plane and red stars indicate the experimental results. 
}

\end{figure}

\textit{Experimental demonstrations.---}
In order to further consolidate the theoretical results, we perform microwave experiment to validate the transverse Kerker effect of a MD emitter. The experimental setup is shown in Fig. 4(a), where a alumina ceramic resonator ($\varepsilon_{r}$ = 9.6, $D$ = 72 mm, $h$ = 35 mm) are used as the dielectric antenna. A home-made loop antennas mimicking the MD emitter is connected to a vector network analyzer (VNA, R\&S ZNB40) to irradiate microwave at the frequencies around 1.90 GHz. An electric monopole antenna connected to the VNA and mounted on the microwave scanning platform (LINBOU, NFS03 Floor Version) is used as an electric field probe to characterize the vectorial electric far-field, which is acquired along a circle trajectory 2$\lambda$ away from the dielectric resonator. Furthermore, microwave absorbers are used to minimize the impact of echo signals. As can be seen from the experimental results shown in Figs. 4(b), the transverse Kerker effect of the loop antenna can be achieved, which also reasonably agrees with the simulation results. 

\begin{figure}[!htb]
\includegraphics[width=\columnwidth]{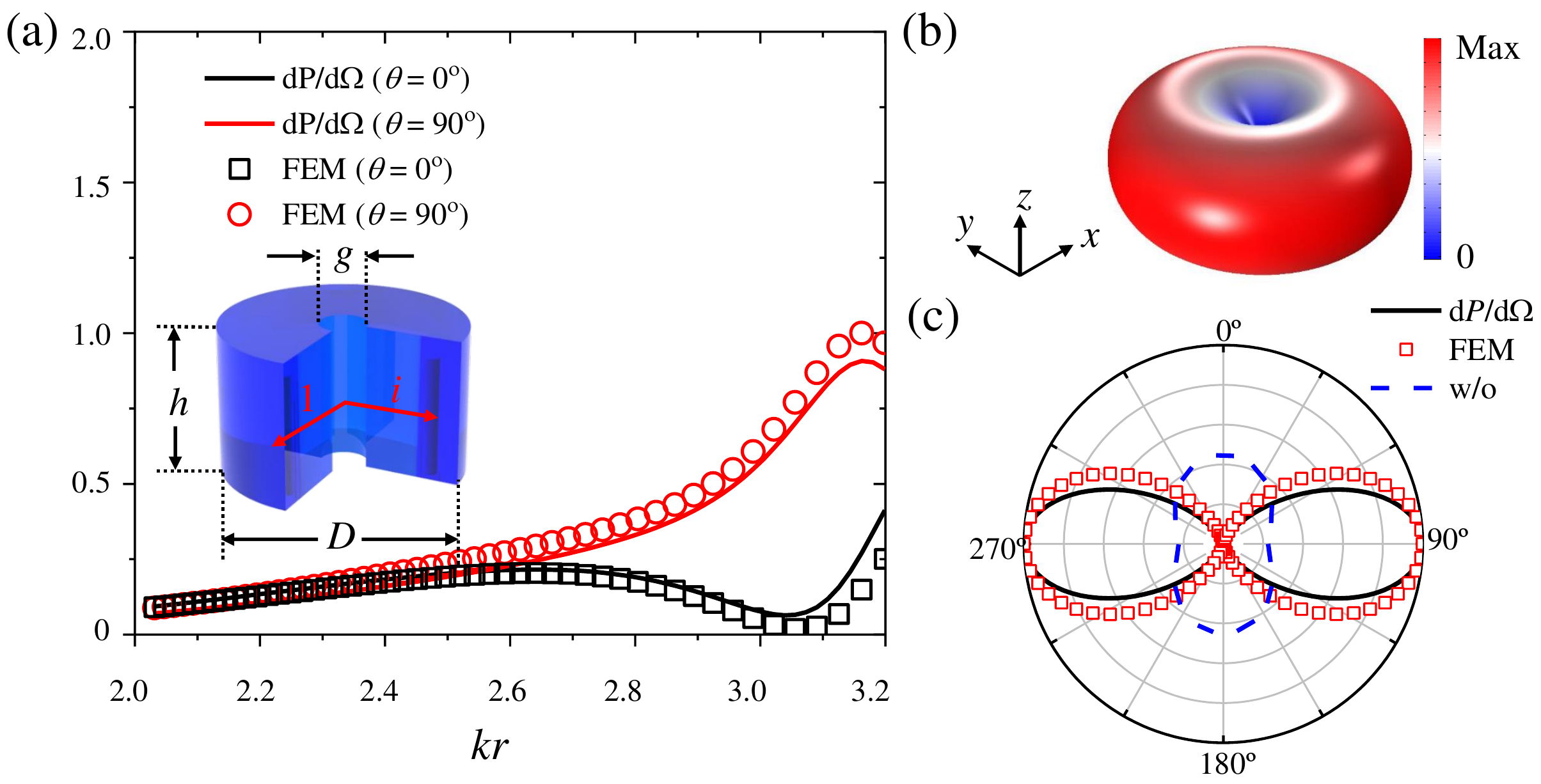}
\caption{\label{fig:fig5} 
(Color online) (a) Normalized time-averaged radiated power as functions of the size parameter $kr$ along two specified directions ($\theta$ = 0$^{\circ}$ and 90$^{\circ}$ when a chiral MD emitter is placed at the center of the hollow dielectric cylinder, as shown in the inset. The result of $\theta$ = 180$^{\circ}$ is the same to the one of $\theta$ = 0$^{\circ}$. (b) and (c) Three- and two-dimensional far-field scattering patterns for the chiral MD emitter interacted with the hollow antenna at the size parameter of $kr$ = 3.04, respectively. The dielectric antenna with a height $h$ = $\lambda_{0}/5.88$ and diameter $D$ =$\lambda_{0}/1.03$ is used where a circular hole with a diameter $g$ = $\lambda_{0}/2.63$ is introduced at the center of the cylinder. The result of far-field radiation patter without the antenna is also provided in (c).
}

\end{figure}

\textit{Discussions}
Although the aforementioned results present the characteristics of transverse KerKer effect of localized electromagnetic sources, it should be pointed out that there is still residual radiation either along the forward or backward direction, which results from the imperfect destructive interference. As mentioned before, the far-field radiation properties are close related to the combined symmetry of the localized electromagnetic sources. Therefore, one can further optimize the transverse Kerker effect of localized electromagnetic source by modifying the combined symmetry of the dipolar emitter and the dielectric antenna. By increasing the symmetry, we show that a hollow dielectric antenna with a perforated hole can be utilized to further suppress the forward and backward radiation of a chiral MD emitter ($\vec{m}_{x}$+$i\vec{m}_{y}$) located at the center of hollow antenna, as indicated in the inset of Fig. 5(a). The normalized time-averaged radiated power along two specified directions ($\theta$ = 0$^{\circ}$ and 90$^{\circ}$) for the dielectric antenna are shown in Fig. 5 (a). The radiative powers for the $-z$ and $+z$ directions are completely coincidence because of symmetry. Notably, the radiation power is 0 at the size parameter $kr$ of 3.04. The corresponding far-field scattering patterns at this condition are shown in Fig. 5 (b) and (c), where the far-field radiation patterns are with two-fold symmetry in $x$-$z$ and $y$-$z$ planes. Furthermore, it should be pointed out that the transverse Kerker effect are general, where the transverse Kerker effect for a chiral ED emitter can also be achieved (see Notes. 3 of the Supplemental Material \cite{Si}).

\textit{Conclusions and outlooks.---}
In summary, we have introduced the concept of \textit{transverse Kerker effect of localized electromagnetic sources} which might provide new prospects to tailor the radiation of complex localized electromagnetic sources. We demonstrate both theoretically and experimentally that a subwavelength dielectric antenna supporting electromagnetic multipolar resonance can be used to realize the transverse Kerker effect of MD and chiral MD/ED emitters where their transverse Kerker conditions are analytically derived. It is envision that new functionalities of radiation management of localized electromagnetic source can be inspired based on the proposed concept. For example, the on-demand steering of radiation direction for single photon source could be achieved utilizing a single and subwavelength dielectric antenna, facilitating the on-chip quantum information process  \cite{jin1,jin2,wang,ren}.

\begin{acknowledgments}
This research was supported by National Key R\&D Program of China (2018YFB1801000), National Natural Science Foundation of China (NSFC) (91750110), the Guangdong Introducing Innovative and Entrepreneurial Teams of "The Pearl River Talent Recruitment Program" (2019ZT08X340), the Research and Development Plan in Key Areas of Guangdong Province (2018B010114002) and the Pearl River Nova Program of Guangzhou (201806010040).
\end{acknowledgments}

\end{document}